\documentclass[10 pt,final,journal,letterpaper,oneside,twocolumn]{IEEEtran}
\usepackage{multicol}   
\usepackage{multirow}
\usepackage[pdftex]{graphicx}
\usepackage[ruled]{algorithm2e}
\usepackage{algorithmic}
\usepackage[small]{caption}
\usepackage{float}
\usepackage{amssymb,amsmath}
\usepackage{graphicx}
\usepackage{subfigure}
\usepackage{xspace}
\usepackage{amsthm}
\usepackage{fancyhdr}

\usepackage{caption,setspace}
\usepackage{amssymb,amsmath}
\usepackage{textcomp} 
\usepackage{epstopdf}
\usepackage{bbding}
\usepackage{cite} 
\usepackage{color}
\usepackage{pifont}

\begin{document}
% Energy-Efficient Communication Design for NOMA LEO Satellite Empowered by Reconfigurable Intelligent Surfaces
% Energy-Efficient RIS-Aided NOMA Communications for LEO Satellite Networks
% Do not put math or special symbols in the title
\title{RIS-Assisted Energy-Efficient LEO Satellite Communications with NOMA}
%Energy Efficiency Optimization in 6G LEO Satellite Networks with RIS-Enabled NOMA Communications
%Energy-Efficient RIS-Enabled NOMA Communication for LEO Satellite Networks
\author{Wali Ullah Khan, \textit{Member, IEEE,} Eva Lagunas, \textit{Senior Member, IEEE,} Asad Mahmood, \\ Symeon Chatzinotas, \textit{Fellow, IEEE,}  Bj\"orn Ottersten, \textit{Fellow, IEEE}  \thanks{This work has been supported by the Luxembourg National Research Fund (FNR) under the project MegaLEO (C20/IS/14767486). An earlier version of this paper is accepted for presentation at the IEEE VTC’2023-Spring, Florence, Italy, in June 2023 \cite{VTC2023}.

Authors are with the Interdisciplinary Centre for Security, Reliability and Trust (SnT), University of Luxembourg
\{waliullah.khan, eva.lagunas, asad.mahmood, symeon.chatzinotas, bjorn.ottersten\}@uni.lu
}}%

\markboth{Submitted to IEEE}%
{Shell \MakeLowercase{\textit{et al.}}: Bare Demo of IEEEtran.cls for IEEE Journals} 
\maketitle
\begin{abstract}
%Low Earth Orbit (LEO) satellite networks are expected to play a crucial role in providing high-speed internet access, reliability and low-latency communication services to ground users worldwide. However, LEO satellite networks face several challenges related to energy and spectral efficiency. In particular, satellites in the LEO network will require a significant amount of power to operate, and this power consumption will increase with the number of satellites in the network. Then, with so many satellites in the sky, there is a risk of spectral congestion, leading to interference and reduced communication quality. Innovative solutions such as energy-efficient technologies and advanced spectrum management techniques will be necessary to address these challenges. Reconfigurable Intelligent surfaces (RIS) and non-orthogonal multiple access (NOMA) have recently received significant attention due to their high energy and spectral efficiency. These technologies have the potential to significantly improve the performance of future 6G LEO satellite networks. With the help of a large number of reconfigurable elements, RIS can improve the signal quality of the ground user and reduce the power consumption of satellite networks. On the other side, NOMA increases the spectral efficiency of the overall network by sharing the same spectrum resources among multiple ground users. 
Low Earth Orbit (LEO) satellite networks are expected to play a crucial role in providing high-speed internet access and low-latency communication worldwide. However, some challenges can affect the performance of LEO satellite networks. For example, they can face energy and spectral efficiency challenges, such as high power consumption and spectral congestion, due to the increasing number of satellites. Furthermore, mobile ground users tend to operate with low directive antennas, which pose significant challenges in closing the LEO-to-ground communication link, especially when operating at a high-frequency range. To overcome these challenges, energy-efficient technologies like reconfigurable intelligent surfaces (RIS) and advanced spectrum management techniques like non-orthogonal multiple access (NOMA) can be employed. RIS can improve signal quality and reduce power consumption, while NOMA can enhance spectral efficiency by sharing the same resources among multiple users. This paper proposes an energy-efficient RIS-assisted downlink NOMA communication for LEO satellite networks. The proposed framework simultaneously optimizes the transmit power of ground terminals of the LEO satellite and the passive beamforming of RIS while ensuring the quality of services. Due to the nature of the considered system and optimization variables, the energy efficiency maximization problem is non-convex. In practice, obtaining the optimal solution for such problems is very challenging. Therefore, we adopt alternating optimization methods to handle the joint optimization in two steps. In step 1, for any given phase shift vector, we calculate satellite transmit power towards each ground terminal using the Lagrangian dual method. Then, in step 2, given the transmit power, we design passive beamforming for RIS by solving the semi-definite programming. We also compare our solution with a benchmark framework having a fixed phase shift design and a conventional NOMA framework without involving RIS. Numerical results show that the proposed optimization framework achieves 21.47\% and 54.9\% higher energy efficiency compared to the benchmark and conventional frameworks. 
\end{abstract}


% For peerreview papers, this IEEEtran command inserts a page break, and

\section{Introduction}
\IEEEPARstart{A}{lthough} 5G networks are still being deployed, researchers are already exploring the potential of 6G technologies, which are expected to offer even faster data speeds, high energy efficiency, lower latency, and more reliable communications \cite{zhang20196g}. To achieve global coverage, these technologies will rely on both terrestrial and non-terrestrial networks, including infrastructure located both on and off the Earth's surface \cite{azari2022evolution}. Low Earth Orbit (LEO) satellite communications, in which satellites orbit at altitudes of 500 to 2000 km, have recently received significant research attention due to their potential to provide worldwide wireless access with low latency compared to Geostationary Earth Orbit and Medium Earth Orbit constellations \cite{su2019broadband}. LEO satellites can be designed to provide high-bandwidth, low-latency communication, which is particularly important for real-time applications \cite{you2020massive}. 

Despite the advantages mentioned above, there are also some potential challenges to consider. For instance, deploying a large number of low LEO satellites will require a significant amount of spectrum and energy resources \cite{kodheli2020satellite}. Furthermore, mobile ground users tend to operate with low directive antennas, which pose significant challenges in closing the LEO-to-ground communication link, especially when operating at a high-frequency range. More specifically, with the existing architecture, mobile users experience a very low data rate. To address this, two possible technologies are non-orthogonal multiple access (NOMA) and Reconfigurable Intelligent Surfaces (RIS). NOMA has the capability to solve the problem of spectrum scarcity in satellite communication by accommodating multiple mobile users over the same spectrum resource simultaneously \cite{gao2021sum}. On the other hand, RIS has the potential to significantly enhance the performance of LEO satellite networks in terms of coverage, energy efficiency, and received signal strength \cite{khan2022opportunities}. RIS consists of multiple sub-wavelength-sized components that may be electronically manipulated to reflect and regulate electromagnetic radiation. 

When applied to LEO satellite networks, RIS can improve data transfer between satellites in orbit and mobile ground users. By strategically installing RIS, we can successfully reflect the incident signals in a desired direction, creating a high-gain antenna \cite{song2023ris}. System capacity, coverage, and energy usage can all improve as a result of improved signal strength and decreased interference. In addition, as LEO satellite networks require a large number of satellites to deliver adequate capacity and global coverage, utilizing RIS can aid in reducing the number of satellites required to get the same level of coverage \cite{ramezani2022toward}. In order to reduce the cost and complexity of establishing and maintaining LEO satellite networks, RIS can improve the efficiency of communication links.

NOMA has recently emerged as a potential multiple-access approach that allows various users to share the same channel on the ground \cite{liu2021spectrum}. According to NOMA, several ground users can share the same frequency band while using varying amounts of power and signal formats to send and receive messages. By maximizing the utilization of the available spectrum, capacity is increased, and spectral efficiency is enhanced \cite{yan2019application}. Combining RIS and NOMA can significantly boost the efficiency of LEO satellites. If RIS is used to create a unique RF environment that is optimized for NOMA, then even more of the available spectrum can be put to good use. To further improve system functionality, NOMA can help minimize interference between satellite and terrestrial networks. Together, RIS and NOMA are two powerful technologies that can greatly benefit LEO satellite operations. 

\subsection{Recent Literature on LEO Satellite Networks}
Several problems with LEO satellite communication networks have been studied by academic and industry researchers. For instance, Li {\em et al.} \cite{li2020hierarchical} have proposed multiple optimization frameworks, including hierarchical, joint and dynamic resource optimizations for maximizing the throughput of integrated satellite-terrestrial networks. The authors of \cite{guo2022efficient} have provided a multi-dimensional resource allocation framework employing weighted greedy and genetic algorithms to improve the performance of LEO satellite networks. Tran {\em et al.} \cite{tran2022satellite} have studied a joint optimization problem in cache-aided LEO satellite networks to maximize the minimum throughput of ground users. The work in \cite{fu2020integrated} has investigated the fairness and reliability of LEO satellite networks by solving the optimization problem using the dual decomposition method. Jia {\em et al.} \cite{jia2021toward} have maximized the total achievable data rate of cooperative terrestrial-satellite networks. Moreover, the paper in \cite{guo2021gateway} has optimized gateway placement by adopting particle swarm optimization to select the best gateway location. Furthermore, Alsharoa {\em et al.} \cite{alsharoa2020improvement} have provided a resource optimization scheme for throughput maximization of satellite-terrestrial networks. In addition, the works in \cite{han2019prediction,zhou2021machine} have also proposed learning-based optimization frameworks to improve the performance of LEO satellite networks.

Besides the above works, some research works have also used NOMA in LEO satellite networks to improve spectral efficiency. In \cite{yan2020delay}, the authors have adopted deep reinforcement learning-based resource optimization for effective capacity maximization in NOMA LEO satellite networks. Gamal {\em et al.} \cite{gamal2022performance} have derived a closed-form expression for outage probability in cooperative NOMA satellite networks. The work in \cite{wang2022adaptive} has proposed a joint resource optimization for minimizing the capacity demand gap in NOMA LEO satellite networks. Zhao {\em et al.} \cite{zhao2022performance} have derived closed-form solutions for the achievable and asymptotic outage probability of satellite networks under hardware impairment and imperfect signal decoding. Moreover, the authors of \cite{yan2021dynamic} have proposed a joint power and channel allocation framework to maximize the sum rate of terrestrial-satellite networks. The research work in \cite{zhao2023multi} has provided a fair resource allocation framework in cooperative satellite networks to maximize the minimum capacity. The authors of \cite{ge2021non} have derived a theoretical expression of ergodic capacity to improve the spectral efficiency of LEO satellite networks. In another paper, the same authors have maximized the sum capacity of NOMA LEO satellite networks by jointly optimizing the user pairing and power allocation \cite{ge2021joint}. Ge {\em et al.} \cite{ge2021performance} have investigated outage probability and ergodic capacity for co-existing NOMA GEO-LEO satellite networks.

Recently, some researchers have integrated RIS with satellite communications. For example, Dong {\em et al.} \cite{dong2021towards} have maximized the weighted sum rate of integrated terrestrial satellite networks by simultaneously optimising the transmit beamforming at the base station and phase shift matrix at RIS. The authors of \cite{zheng2022ris} have provided a new framework to enhance the average throughput of RIS-enabled LEO networks by jointly optimizing the orientation and passive beamforming at RIS. Lee {\em et al.} \cite{lee2021performance} have jointly optimized the active and passive beamforming to maximize the received signal-to-noise ratio in RIS-assisted LEO satellite networks. Moreover, Khan {\em et al.} \cite{khan2022ris} have proposed a sustainable framework for maximizing the spectral efficiency of RIS-assisted GEO satellite networks by jointly optimizing the transmit power of the satellite and phase shift at RIS. The work in \cite{zheng2022intelligent} has designed a new architecture to enhance the overall channel gain in RIS-assisted LEO satellite networks by jointly optimising the transmit and receive beamforming. Further, Tekbiyik {\em et al.} \cite{tekbiyik2022reconfigurable} have investigated bit error rate and achievable rate for RIS-assisted terahertz communication in LEO satellite networks. Xu {\em et al.} \cite{xu2021intelligent} have also studied a physical layer security problem in RIS-assisted cognitive radio satellite-terrestrial integrated networks. Of late, the author of \cite{dong2021towards} have extended their work in \cite{dong2022intelligent} to maximize the weighted sum rate of RIS-assisted LEO satellite networks by jointly optimizing user scheduling, transmit beamforming and phase shift design.
%%%%%%%%%%%%%%%%
\begin{figure*}[!t]
\centering
\includegraphics [width=0.55\textwidth]{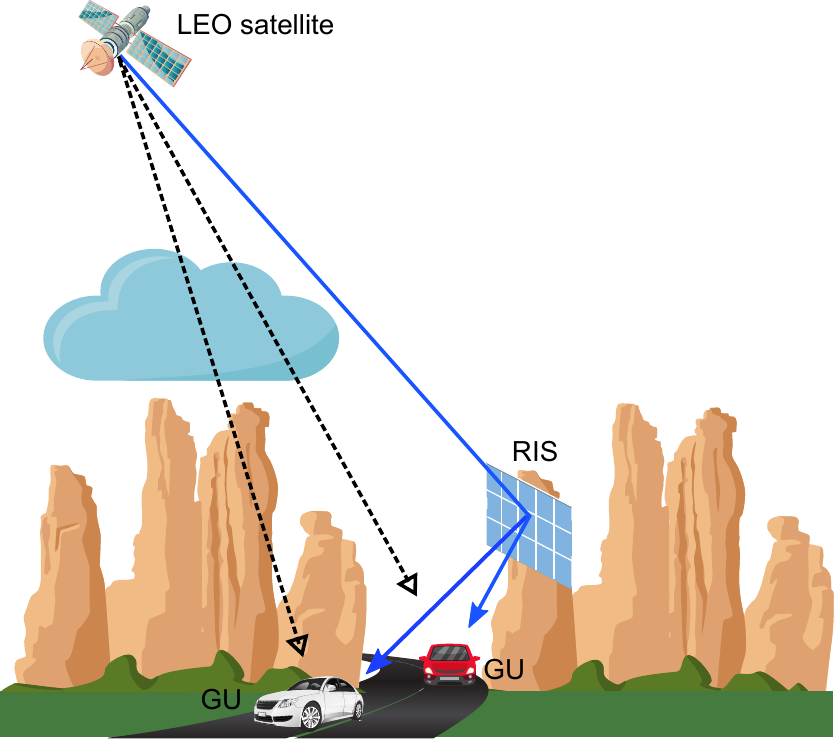}
\caption{System model of RIS-assisted LEO satellite communication with NOMA}
\label{blocky}
\end{figure*}
%%%%%%%%%%%%%
\subsection{Motivation and Contributions}
Although extensive work has been done on LEO satellite communications, there still exists some open gaps that need to be explored. For example, the research works \cite{li2020hierarchical,guo2022efficient,tran2022satellite,fu2020integrated,jia2021toward,guo2021gateway,alsharoa2020improvement,han2019prediction,zhou2021machine} have studied LEO satellite networks by adopting orthogonal multiple access (OMA) techniques. However, they do not consider NOMA and RIS in their proposed models. Then, some works proposed NOMA in LEO satellite communication networks \cite{yan2020delay,gamal2022performance,wang2022adaptive,zhao2022performance,yan2021dynamic,zhao2023multi,ge2021non,ge2021joint,ge2021performance}, but their system models did not consider RIS technology. Of late, researchers have integrated RIS into LEO satellite networks using conventional OMA techniques \cite{dong2021towards,zheng2022ris,lee2021performance,khan2022ris,zheng2022intelligent,tekbiyik2022reconfigurable,xu2021intelligent,dong2022intelligent}. However, NOMA is not considered as a multiple access technique. To the best of our knowledge, no work exists on RIS-assisted NOMA LEO satellite communication; hence, it is an open topic to study. Therefore, this paper proposes an optimization framework for RIS-assisted NOMA LEO satellite communication. In particular, we maximize the achievable energy efficiency of the system by simultaneously optimizing the transmit power of the LEO satellite and passive beamforming at RIS while ensuring the quality of services. The main contribution of this work can be summarized as follows.
\begin{enumerate}
    \item This paper considers the RIS-assisted NOMA LEO satellite communication network, where a satellite serves ground mobile users using NOMA protocol. We consider the ground users are mobile and direct links from satellite to ground users experience poor link budget. To overcome this issue and improve the received channel gains, we consider that a RIS system is mounted on a strategic place to assist the satellite signal to ground users. Therefore, ground users receive signals through direct and RIS-assisted links. The aim of this work is to efficiently allocate system resources and improve the total achievable energy efficiency of the satellite network. An energy efficiency maximization problem is formulated to simultaneously optimizes the transmit power of the satellite and passive beamforming of the RIS while ensuring the quality of services of ground users. 
    \item The formulated optimization problem is non-convex due to the interference term in objective function and quality of services constraints. The problem is also coupled with two optimization variables, and the objective function is fractional. Therefore, achieving a joint optimal solution is very challenging for the original problem. To reduce the complexity and make the optimization problem more tractable, we adopt an alternating optimization approach. According to this approach, the optimization problem can be efficiently solved in two steps without compromising the generality of the original problem. In the first step, the transmit power of the satellite can be calculated for any given beamforming at the RIS system. Then, in the second step, the passive beamforming at the RIS system can be designed given the optimal power of the satellite.
    \item For the given beamforming at the RIS, the problem can be simplified as a power allocation at the LEO satellite. To solve the power allocation problem, we first exploit the Dinkelbach method to transform the fractional objective into a non-fractional function. Then, we employ the successive convex approximation and Lagrangian dual method to obtain an efficient solution. After having optimal power allocation, the passive beamforming problem at RIS can be further simplified. Since the direct link from satellite to ground users have no impact on the passive beamforming at RIS, it can be safely ignored. To handle this problem, we perform some straightforward transformations to make the problem semi-definite programming, which is convex in nature. This convex optimization problem can be efficiently solved by any standard solver such as CVX.  
    \item To validate the proposed optimization solution, numerical results are provided. For evaluation of the performance of the proposed framework, two benchmark NOMA optimization frameworks are also proposed. In the first benchmark framework, we only optimize the transmit power of the LEO satellite while the phase shift at the RIS system is considered fixed. In the second benchmark optimization framework, we consider a conventional NOMA satellite network without taking the RIS system. The results demonstrate the benefits of the proposed optimization framework by outperforming the other benchmark frameworks. Moreover, the proposed solution converges within a few iterations, which shows low complexity.
\end{enumerate}

The remainder of the paper is structured as follows. Section II studies the system model of RIS-assisted NOMA LEO satellite communication. Section III provided the proposed energy-efficient solution. Section IV provides numerical results and discussion, while Section V concludes this paper.

\section{System Model and Problem Formulation}
Consider an LEO satellite that communicates with mobile ground users (GUs) in mountain areas using the NOMA protocol, as shown in Fig. \ref{blocky}. The LEO satellite uses Ka-band and multi-beam technology. Although NOMA has the potential to accommodate multiple users over the same resource block, we assume that the LEO satellite accommodates only two GUs at any given time to reduce the
computational complexity and delay incurred at GUs due to SIC decoding process \cite{7934461}. However, the proposed model will be generalized in future studies to accommodate the case of serving multiple GUs on a single resource block. We consider that the GUs are mobile and may experience large-scale fading, resulting in performance degradation. Therefore, we also consider using RIS, strategically mounted on the mountain, to assist with the signal delivery of NOMA GUs from the LEO satellite \cite{10003076}. It is assumed that the GUs are equipped with one antenna, RIS consists of multiple reconfigurable elements, and the channel state information in the entire system is available.

Let us denote the LEO satellite as $l$ and the mobile GUs as GU $i$ and GU $j$, respectively. Moreover, the reconfigurable elements are defined as $\mathcal M=\{m|1,2,3,\dots,M\}$, where $m$ indexes a single element. Accordingly, the $M\times M$ diagonal matrix of RIS can be defined as $\boldsymbol\Theta=\text{diag}\{\alpha_1,\alpha_2,\alpha_3,\dots, \alpha_M\}$, where $\alpha_m$ denotes the reconfigurable coefficent of element $m$ and satisfy $|\alpha_m|=1, \ \forall m\in \{1,2,3,\dots, M\}$. Due to different geographical locations, we assume that the channel gains of GUs are not similar. Thus, the direct channel gains from LEO satellite to GUs can be sorted as $h_{l, i}\geq h_{l,j}$ without loss of generality. By considering the block faded channel model, $h_{l,\iota}\in\{i,j\}$ can be expressed as
\begin{align}
h_{l,\iota}=\hat{h}_{l,\iota}e^{\hat j\pi\zeta_{l,\iota}},
\end{align}
where $\hat{h}_{l,\iota}$ denotes the complex-valued channel gain, $\zeta_{l,\iota}$ is the Doppler shift, and $\hat j=\sqrt{-1}$. Considering both small and large-scale fading, the complex-valued channel gain $\hat{h}_{l,\iota}$ can be further stated as 
\begin{align}
\small
\hat{h}_{l,\iota}=\sqrt{G_lG_{l,\iota} \Big(\frac{c}{4\pi f_c d_{l,\iota}}\Big)^2}
\end{align}
where $c$ represents the speed of light, $f_c$ is the carrier frequency, $d_{l,\iota}$ denotes the distance from LEO satellite to GU, $G_{l,\iota}$ shows the received antenna gain, and $G_{l}$ denotes the transmit antenna gain, respectively. Note that the transmit antenna gain mostly depends on radiation pattern and GU geographical location \cite{ghosh20195g}, which can be approximated as
\begin{align}
 G_{l} =   G_{max}\left[\frac{J_1(\vartheta_{l,\iota})}{2\vartheta_{l,\iota}}+36\frac{J_3(\vartheta_{l,\iota})}{\vartheta^3_{l,\iota}}\right]^2,
\end{align}
where $G_{max}$ represents the maximum gain observed at the beam centre, $\vartheta_{l,\iota}=2.07123\sin{(\theta_{l,\iota})}/\sin(\theta_{3dB})$ such that $\theta_{l,\iota}$ denotes the angle between GU and the centre of LEO satellite beam for a given GU location. The angle of 3 dB loss related to the beam's centre is given by $\theta_{3dB}$. Moreover, $J_1$ and $J_3$ are the first-kind Bessel functions having orders 1 and 2. Let us assume that $x_l$ is the superimposed signal of LEO satellite for GU $i$ and GU $j$, such as
\begin{align}
x_l=\sqrt{P_l\varrho_{l,i}}x_{l,i}+\sqrt{P_l\varrho_{l,j}}x_{l,j},
\end{align}
where $P_l$ is the transmit power of LEO satellite while $\varrho_{l,i}$ and $\varrho_{l,j}$ are the power allocation coefficients. Further, $x_{l,i}$ and $x_{l,j}$ are the unit power signals. The signals that GU $i$ and GU $j$ receive from the LEO satellite can be written as
%%%%%%%%%%%%%%%%%%%%%%%%%%%%%%
\begin{align}
y_{l,i}&=(h_{l,i}+\boldsymbol{g}_{l,m}\boldsymbol{\Theta}\boldsymbol{f}_{m,i})x_l+\omega_{l,i},\label{y1}
\\
y_{l,j}&=(h_{l,j}+\boldsymbol{g}_{l,m}\boldsymbol{\Theta}\boldsymbol{f}_{m,j})x_l+\omega_{l,j},\label{y2}
\end{align}
%%%%%%%%%%%%%%%%%%%%%%%%%%%%%
where $\omega_{l,\iota}$ is the additive white Gaussian noise (AWGN), $\boldsymbol{g}^H_{l,m}\in M\times1$ denotes the channel gains from LEO satellite to RIS. Moreover, $\boldsymbol{f}_{m,\iota}=\hat{f}_{m,\iota}d^{-\beta/2}_{l,\iota}\in M\times1$ represent the channel gains from RIS to GU $\iota$, where $\hat{f}_{m,\iota}$ is the channel coefficient, $d_{l,\iota}$ is the distance from RIS to GUs, and $\beta$ denotes the pathloss exponent.
Following the downlink NOMA protocol, GU with a higher received channel gain can apply SIC decoding technique to subtract the signal of another user before decoding the desired signal. However, the GU with a lower received channel gain cannot apply SIC and decode the desired signal by treating the signal of other GU as noise. Based on this observation, the signal-to-noise ratio of GU $i$ can be expressed as 
\begin{align}
\small
\gamma_{l,i}=P_l\varrho_{l,i}|h_{l,i}+\boldsymbol{g}_{l,m}\boldsymbol{\Theta}\boldsymbol{f}_{m,i}|^2/\sigma^2,\label{7}
\end{align}
where $\sigma^2$ is the variance of AWGN. Accordingly, the signal-to-interference plus noise ratio of GU $j$ can be written as
\begin{align}
\gamma_{l,j}=\frac{P_l\varrho_{l,j}|h_{l,j}+\boldsymbol{g}_{l,m}\boldsymbol{\Theta}\boldsymbol{f}_{m,j}|^2}{\sigma^2+P_l\varrho_{l,i}|h_{l,j}+\boldsymbol{g}_{l,m}\boldsymbol{\Theta}\boldsymbol{f}_{m,j}|^2},\label{8}
\end{align}
where the second term in the denominator is the interference of GU $i$. Now we can define the data rate of GU $i$ and GU $j$ from the LEO satellite as
\begin{align}
R_{l,i} = \log_2(1+\gamma_{l,i}),\label{9}
\end{align}
\begin{align}
R_{l,j} = \log_2(1+\gamma_{l,j}),\label{10}
\end{align}

%%%%%%%%%%%%%%%%%%%%%%%%%%%%%%%%%%%%%%%%%%
\begin{figure*}[!t]
\begin{align}
&\mathcal L(\varrho_{l,i},\varrho_{l,j},\boldsymbol{\lambda}) = (\Psi_{l,i}\log_2(\gamma_{l,i})+\Omega_{l,i}+\Psi_{l,j}\log_2(\gamma_{l,j})+\Omega_{l,j})-\phi^{t-1}(P_l(\varrho_{l,i+\varrho_{l,j}})+p_c)\nonumber\\&+\lambda_{1}\left(P_l\varrho_{l,i}|h_{l,i}+\boldsymbol{g}_{l,m}\boldsymbol{\Theta}\boldsymbol{f}_{m,i}|^2-\gamma_{min}(\sigma^2)\right)+\lambda_{2}(P_l\varrho_{l,j}|h_{l,j}+\boldsymbol{g}_{l,m}\boldsymbol{\Theta}\boldsymbol{f}_{m,j}|^2-\gamma_{min}(\sigma^2\nonumber\\&+P_l\varrho_{l,i}|h_{l,j}+\boldsymbol{g}_{l,m}\boldsymbol{\Theta}\boldsymbol{f}_{m,j}|^2)) +\lambda_3(P_T-P_l\varDelta )+\lambda_4(1-(\varrho_{l,i}+\varrho_{l,j})), \tag{15}\label{250}
\end{align}\hrulefill
\begin{align}
&\dfrac{L(\varrho_{l,i},\varrho_{l,j},\boldsymbol{\lambda})}{\partial \varrho_{l,i}}=\varrho_{l,i}^2 \Big(-O_{l,j} P_l (-\lambda_1 O_{l,i} P_l+(\lambda_4+P_l(\lambda_3+\phi^{t-1}+\lambda_2 O_{l,j}\gamma_{min}))\sigma^2)\Big)+\varrho_{l,i}\Big(-\sigma(-\lambda_1 O_{l,i} P_l\nonumber\\& +(\lambda_4 +(\lambda_3+\phi^{t-1})P_l)\sigma^2+O_{l,j} P_l (-\Psi_{l,i}+\Psi_{l,j}+\lambda_2\gamma_{min}\sigma^2))\Big) +\Psi_{l,i}\sigma^2=0. \tag{17}\label{251}
\end{align}\hrulefill
\end{figure*}
The objective of this work is to maximize the achievable energy efficiency of the NOMA LEO satellite system. In this work, the total energy efficiency can be defined as the total achievable capacity divided by the system's total power consumption. This objective can be achieved by optimising the transmit power of the LEO satellite and passive beamforming of RIS subject to the quality of services constraints. The optimization problem can be formulated as
%%%%%%%%%%%%%%%%%%%%%%%%%%%
\begin{alignat}{2}
&\underset{{\boldsymbol{\varrho}, \boldsymbol{\Theta}}}{\text{max}} \ \frac{\log_2(1+\gamma_{l,i})+\log_2(1+\gamma_{l,j})}{P_l(\varrho_{l,i+\varrho_{l,j}})+p_c}\label{OP} \\
 s.t.\ &    \gamma_{l,i}\geq \gamma_{min}, \tag{11a}\label{11a}\\
&  \gamma_{l,j}\geq \gamma_{min},\tag{11b} \label{11b}\\
& 0\leq P_l(\varrho_{l,i}+\varrho_{l,j})\leq P_T,\tag{11c} \label{11c}\\
& |\alpha_m|=1,\ \forall m\in M,\tag{11d} \label{11d}\\
& \varrho_{l,i}+\varrho_{l,j}\leq 1,\tag{11e} \label{11e}\\
 & 0\leq\varrho_{l,i}\leq1, 0\leq\varrho_{l,j}\leq1,\tag{11f} \label{11f}
\end{alignat}
%%%%%%%%%%%%%%%%%%%%%%%%%%
where the objective in (\ref{OP}) is to maximize the achievable energy efficiency of the RIS-assisted NOMA LEO satellite network. The term $p_c$ denotes the circuit power consumption of the system. The constraints (\ref{11a}) and (\ref{11b}) guarantee the quality of services for GU $j$ and GU $j$. Constraint (\ref{11c}) controls the power transmission of the LEO satellite. Constraint (\ref{11d}) invokes the phase shift of the RIS system. Constraints (\ref{11e}) and (\ref{11f}) allocate power to GU $i$ and GU $j$ based on downlink NOMA. 

\section{Proposed Energy-Efficient Solution}
The optimization problem in (\ref{OP}) is non-convex due to several factors, i.e., it couples with two variables, interference in rate expression and the fractional objective function. Thus, obtaining a joint optimal solution is challenging because of its high complexity. To efficiently solve this problem, we exploit an alternating optimization approach. Based on this method, the optimization problem in (\ref{OP}) can be efficiently solved in two steps. In the first step, the transmit power of GU $i$ and GU $j$ can be calculated at the LEO satellite for any given passive beamforming at RIS. Then in the second step, the passive beamforming at RIS is computed given the transmit power at the LEO satellite.
%%%%%%%%%%%%%%%%%%%%%%%%%%%%%%%%%%%%%%%%%%%%%%%%%%%%%%%%%%%%%%%%%%%%
\begin{figure*}[!t]
\begin{align}
A&=\sigma^2 ((-\lambda_1 O_{l,i} P_l+(\lambda_4+(\lambda_3+\phi^{t-1})\sigma^2)+O_{l,j} P_l(-\Psi_{l,i}+\Psi_{l,j}+\lambda_2 \gamma_{min}\sigma^2)),\nonumber\\
B&=\sigma^4(4 O_{l,j} P_l \Psi_{l,i}(-\lambda_1 O_{l,i} P_l+(\lambda_4+P_l(\lambda_3+\phi^{t-1}+\lambda_2 O_{l,j} \gamma_{min}))\sigma^2)\nonumber\\&+((-\lambda_1 O_{l,i} P_l+(\lambda_4+(\lambda_3+\phi^{t-1})P_l)\sigma^2)+O_{l,j} P_l(-\Psi_{l,i}+\Psi_{l,j}+\lambda_2\gamma_{min}\sigma^2))^2),\tag{19}\label{252}\\
C&=-2 O_{l,j} P_l (-\lambda_1 O_{l,i} P_l+(\lambda_4+P_l (\lambda_3+\phi^{t-1}+\lambda_2O_{l,j} \gamma_{min}))\sigma^2).\nonumber 
\end{align}\hrulefill
\end{figure*}
%%%%%%%%%%%%%%%%%%%%%%%%%%%%%%%%%%%%%%%%%%
\subsection{NOMA Power Allocation at LEO Satellite: Step-1}
For any given passive beamforming $\boldsymbol{\Theta}$ at RIS, the problem in (\ref{OP}) can be simplified as power allocation optimization at LEO satellite, such as
%%%%%%%%%%%%%%%%%%%%%%%%%%%
\begin{alignat}{2}
&\underset{\varrho_{l,i},\varrho_{l,j}}{\text{max}}\ \frac{\log_2(1+\gamma_{l,i})+\log_2(1+\gamma_{l,j})}{P_l(\varrho_{l,i+\varrho_{l,j}})+p_c}\label{OP1} \\
 s.t.\ & P_l\varrho_{l,i}|h_{l,i}+\boldsymbol{g}_{l,m}\boldsymbol{\Theta}\boldsymbol{f}_{m,i}|^2\geq \gamma_{min} (\sigma^2), \tag{12a}\label{12a}\\
 & P_l\varrho_{l,j}|h_{l,j}+\boldsymbol{g}_{l,m}\boldsymbol{\Theta}\boldsymbol{f}_{m,j}|^2\geq \gamma_{min} (\sigma^2+P_l\varrho_{l,i}\nonumber\\&|h_{l,j}+\boldsymbol{g}_{l,m}\boldsymbol{\Theta}\boldsymbol{f}_{m,j}|^2),\ (\ref{11c}), (\ref{11e}).\tag{12b}\label{12b}
\end{alignat}
%%%%%%%%%%%%%%%%%%%%%%%%%%
The proposed optimization in (\ref{OP1}) is a fractional problem due to the objective function. It can be efficiently transformed into a non-fractional, which can be expressed as
%%%%%%%%%%%%%%%%%%%%%%%%%%%
\begin{alignat}{2}
&\underset{\varrho_{l,i},\varrho_{l,j}}{\text{max}}\ \log_2(1+\gamma_{l,i})+\log_2(1+\gamma_{l,j})\nonumber\\&-\phi^{t-1}(P_l(\varrho_{l,i+\varrho_{l,j}})+p_c)\label{OP12} \\
 s.t.\ & (\ref{12a}),(\ref{12b}), (\ref{11c}),(\ref{11e}), \nonumber
\end{alignat}
where $\phi$ denotes a non-negative parameter and $t$ indexes iteration number. To solve problem (\ref{OP12}), the $\boldsymbol{\gamma}$ and $\boldsymbol{\varrho}$ can be updated in each iteration as $\phi^t=(\log_2(1+\gamma_{l,i})+\log_2(1+\gamma_{l,j}))/(P_l(\varrho_{l,i}+\varrho_{l,j})+p_c)$. Moreover, the maximum energy efficiency in (\ref{OP12}) can be computed as $\eta=(\log_2(1+\gamma^t_{l,i})+\log_2(1+\gamma^t_{l,j}))-\phi^{t-1}(P_l(\varrho^t_{l,i}+\varrho^t_{l,j})+p_c)$. During the iterative process, the value of $\phi$ increases while the value of $\eta$ reduces and approaches to zero. The maximum energy efficiency is achieved when $\phi$ is maximum and $\eta=0$.

Next, it is important to see how the optimization in (\ref{OP12}) can be solved for a given $\phi$. It is clear that the problem (\ref{OP12}) is non-convex due to the non-concave objective function. 
To simplify the problem and make it more manageable, we use the successive convex approximation method.
%The basic idea of the SCA method is to transform the non-convex optimization problem into a sequence of convex optimization subproblems, where each problem approximates the original optimization problem. 
By applying this method, the sum capacity in the objective of the problem (\ref{OP12}) can be expressed as
$\Psi_{l,i}\log_2(\gamma_{l,i})+\Omega_{l,i}+\Psi_{l,j}\log_2(\gamma_{l,j})+\Omega_{l,j}$ with
the values of
$\Psi_{l,\iota}=\frac{\gamma_{l,\iota}}{1+\gamma_{l,\iota}}$ and $\Omega_{l,\iota}=\log_2(1+\gamma_{l,\iota})-\Psi_{l,\iota}\log_2(\gamma_{l,\iota}),\ \iota\in\{i,j\}$.
Now the optimization problem in (\ref{OP12}) can be reformulated as
%%%%%%%%%%%%%%%%%%%%%%%%%%%
\begin{alignat}{2}
&\underset{\varrho_{l,i},\varrho_{l,j}}{\text{max}}(\Psi_{l,i}\log_2(\gamma_{l,i})+\Omega_{l,i}+\Psi_{l,j}\log_2(\gamma_{l,j})+\Omega_{l,j})\nonumber\\&-\phi^{t-1}(P_l(\varrho_{l,i+\varrho_{l,j}})+p_c)\label{OP13} \\
 s.t.\ & (\ref{12a}),(\ref{12b}), (\ref{11c}),(\ref{11e}), \nonumber
\end{alignat}
Now we employ the Lagrangian dual method to efficiently solve the optimization problem (\ref{OP13}). The Lagrangian function to solve problem (\ref{OP13}) can be defined as Equation (\ref{250}) on the top of this page,
%\begin{align}
%&\mathcal L(\varrho_{l,i},\varrho_{l,j},\boldsymbol{\lambda}) = (\Psi_{l,i}\log_2(\gamma_{l,i})+\Omega_{l,i}\nonumber\\&+\Psi_{l,j}\log_2(\gamma_{l,j})+\Omega_{l,j})-\phi^{t-1}(P_l(\varrho_{l,i+\varrho_{l,j}})+p_c)\nonumber\\&+\lambda_{1}\left(P_l\varrho_{l,i}|h_{l,i}+\boldsymbol{g}_{l,m}\boldsymbol{\Theta}\boldsymbol{f}_{m,i}|^2-\gamma_{min}(\sigma^2)\right)\nonumber\\ &+\lambda_{2}(P_l\varrho_{l,j}|h_{l,j}+\boldsymbol{g}_{l,m}\boldsymbol{\Theta}\boldsymbol{f}_{m,j}|^2-\gamma_{min}(\sigma^2\nonumber\\&+P_l\varrho_{l,i}|h_{l,j}+\boldsymbol{g}_{l,m}\boldsymbol{\Theta}\boldsymbol{f}_{m,j}|^2)) +\lambda_3(P_T-P_l\varDelta )\nonumber\\&+\lambda_4(1-(\varrho_{l,i}+\varrho_{l,j})), \label{250}
%\end{align}
where $\boldsymbol{\lambda}=\{\lambda_1,\lambda_2,\lambda_3,\lambda_4\}$ and $\varDelta=\varrho_{l,i}+\varrho_{l,j}$. Now we use Karush-Khun Tucker (KKT) conditions to (\ref{250}) such as
\begin{align}
\frac{\partial \mathcal L(\varrho_{l,i},\varrho_{l,j},\boldsymbol{\lambda})}{\partial \varrho_{l,i},\varrho_{l,j}}|_{\varrho_{l,i},\varrho_{l,j}=\varrho_{l,i}^*,\varrho_{l,j}^*}=0,\tag{16}
\end{align}
After the straightforward derivative of (\ref{250}), we obtain a polynomial, which can be written as Equation (\ref{251}) on the top of this page.
%\begin{align}
 %   &\dfrac{L(\varrho_{l,i},\varrho_{l,j},\boldsymbol{\lambda})}{\partial \varrho_{l,i}}=\varrho_{l,i}^2 \Big(-O_{l,j} P_l (-\lambda_1 O_{l,i} P_l+(\lambda_4+P_l\nonumber\\&(\lambda_3+\phi^{t-1}+\lambda_2 O_{l,j}\gamma_{min}))\sigma^2)\Big)+\varrho_{l,i}\Big(-\sigma(-\lambda_1 O_{l,i} P_l\nonumber\\& +(\lambda_4 +(\lambda_3+\phi^{t-1})P_l)\sigma^2+O_{l,j} P_l (-\Psi_{l,i}+\Psi_{l,j}+\lambda_2\nonumber\\&\gamma_{min}\sigma^2))\Big) +\Psi_{l,i}\sigma^2=0
%\end{align}
Now solving $\varrho_{l,i}$, it can be expressed as
\begin{align}
 \varrho^*_{l,i}=\dfrac{A \pm \sqrt{B}}{C}  \tag{18} \label{18}
\end{align}
where the values of $A$, $B$, and $C$ can be found in Equation (\ref{252}) on the top of the next page.
%\begin{align}
%A=&\sigma^2 ((-\lambda_1 O_{l,i} P_l+(\lambda_4+(\lambda_3+\phi^{t-1})\sigma^2)+O_{l,j} P_l\nonumber\\& (-\Psi_{l,i}+\Psi_{l,j}+\lambda_2 \gamma_{min}\sigma^2)),\\
%B=&\sigma^4(4 O_{l,j} P_l \Psi_{l,i}(-\lambda_1 O_{l,i} P_l+(\lambda_4+P_l(\lambda_3+\phi^{t-1}+\nonumber\\&\lambda_2 O_{l,j} \gamma_{min}))\sigma^2)+((-\lambda_1 O_{l,1} P_l+(\lambda_4+(\lambda_3+\phi^{t-1})\nonumber\\&P_l)\sigma^2)+O_{l,j} P_l(-\Psi_{l,i}+\Psi_{l,j}+\lambda_2\gamma_{min}\sigma^2))^2),\\
%C=&-2 O_{l,j} P_l (-\lambda_1 O_{l,i} P_l+(\lambda_4+P_l (\lambda_3+\phi^{t-1}+\lambda_2 \nonumber\\&O_{l,j} \gamma_{min}))\sigma^2)
%\end{align}
In addition, the values of $O_{l,i}$ and $O_{l,j}$ can be given as
\begin{align}
&O_{l,i}=|h_{l,i}+\boldsymbol{g}_{l,m}\boldsymbol{\Theta}\boldsymbol{f}_{m,i}|^2.\tag{20}\\
&O_{l,j}=|h_{l,j}+\boldsymbol{g}_{l,m}\boldsymbol{\Theta}\boldsymbol{f}_{m,j}|^2.\tag{21}
\end{align}
After computing $\varrho^*_{l,i}$, the value of $\varrho^*_{l,j}$ can be efficiently achieve as
\begin{align}
\varrho^*_{l,j}=1-\varrho^*_{l,i}. \tag{22}\label{22}
\end{align}
Subsequently, vector $\boldsymbol{\lambda}$ can be iteratively updated using the sub-gradient method as
\begin{align}
\lambda_1(t+1)&=[\lambda_1(t)+\mu(t)\times(P_l\varrho_{l,i}|h_{l,i}\nonumber\\&+\boldsymbol{g}_{l,m}\boldsymbol{\Theta}\boldsymbol{f}_{m,i}|^2-\gamma_{min}(\sigma^2))]^+,\tag{23}\label{23}
\end{align}
\begin{align}
\lambda_2(t+1)&=[\lambda_1(t)+\mu(t)\times(P_l\varrho_{l,j}|h_{l,j}+\boldsymbol{g}_{l,m}\boldsymbol{\Theta}\boldsymbol{f}_{m,j}|^2\nonumber\\&-\gamma_{min}(\sigma^2+P_l\varrho_{l,i}|h_{l,j}+\boldsymbol{g}_{l,m}\boldsymbol{\Theta}\boldsymbol{f}_{m,j}|^2))]^+,\tag{24}\label{24}
\end{align}
\begin{align}
\lambda_3(t+1)=[\lambda_3(t)+\mu(t)\times(P_T-P_l(\varrho_{l,i}+\varrho_{l,j}))]^+,\tag{25}\label{25}
\end{align}
\begin{align}
\lambda_4(t+1)=[\lambda_3(t)+\mu(t)\times(1-(\varrho_{l,i}+\varrho_{l,j}))]^+,\tag{26}\label{26}
\end{align}
where $\mu$ denotes the non-negative step size. In each iteration, the Lagrangian and optimization variables are updated. The iterative process stops after convergence.
%Finally, we update the vector of Lagrangian variables as
%\begin{align}
%\lambda_1(t+1)=\left[\lambda_1(t)+\varepsilon(t)\times \left(\gamma_{l,i}-2^{\frac{2R_{min}-\Psi_{l,i}}{\Omega_{l,i}}} \right)\right]^+,
%\end{align}
%\begin{align}
%\lambda_2(t+1)=\left[\lambda_2(t)+\varepsilon(t)\times \left(\gamma_{l,j}-2^{\frac{2R_{min}-\Psi_{l,j}}{\Omega_{l,j}}} \right)\right]^+,
%\end{align}
%\begin{align}
%\lambda_3(t+1)=\left[\lambda_3(t)+\varepsilon(t)\times \left(P_t-P_l(\varrho_l,i+\varrho_{l,j}) \right)\right]^+,
%\end{align}
%\begin{align}
%\lambda_4(t+1)=[\lambda_4(t)+\varepsilon(t)\times \left(1-(\varrho_l,i+\varrho_{l,j}) \right)]^+,
%\end{align}
%where $[.]^+=\max\{0,.\}$, $t$ indexes iteration number and $\varepsilon(t)$ is the non-negative step size. During the optimization process, the values of all variables in each iteration can be updated, and the iterative process will continue until convergence.

%%%%%%%%%%%%%%%%%%%%%%%%%%%%%%%%%%%%%%%%%%
\begin{figure*}[!t]
\begin{align}
&\log_2(1+\text{Tr}(\boldsymbol{\Xi}\boldsymbol{G}_{l,i})/\sigma^2)+\log_2\bigg(1+\frac{\text{Tr}(\boldsymbol{\Xi}\boldsymbol{G}_{l,j})}{\text{Tr}(\boldsymbol{\Xi}\boldsymbol{\bar{G}}_{l,j})+\sigma^2}\bigg)=(\log_2(\sigma^2+\text{Tr}(\boldsymbol{\Xi}\boldsymbol{G}_{l,i}))-\log_2(\sigma^2))\nonumber\\
&+(\log_2(\text{Tr}(\boldsymbol{\Xi}\boldsymbol{\bar{G}}_{l,j})+\sigma^2+\text{Tr}(\boldsymbol{\Xi}\boldsymbol{G}_{l,j}))-\log_2(\text{Tr}(\boldsymbol{\Xi}\boldsymbol{\bar{G}}_{l,j})+\sigma^2)). \tag{34}\label{253}
\end{align}\hrulefill
\end{figure*}
%%%%%%%%%%%%%%%%%%%%%%%%%%%%%%%%%%%%%%%%%%

\subsection{Passive Beamforming at RIS: Step-2}
Under given $\varrho^*_{l,i},\varrho^*_{l,j}$ at LEO satellite, the passive beamforming at RIS can be further simplified. We can observe that the direct links from satellite to GUs have no impact on the passive beamforming. Thus it can be efficiently ignored. The Equations (\ref{7}) and (\ref{8}) can be then reformulated as
\begin{align}
\bar\gamma_{l,i}=P_l\varrho_{l,i}|\boldsymbol{g}_{l,m}\boldsymbol{\Theta}\boldsymbol{f}_{m,i}|^2/\sigma^2,\tag{27}\label{27}\\
\bar\gamma_{l,j}=\frac{P_l\varrho_{l,j}|\boldsymbol{g}_{l,m}\boldsymbol{\Theta}\boldsymbol{f}_{m,j}|^2}{\sigma^2+P_l\varrho_{l,i}|\boldsymbol{g}_{l,m}\boldsymbol{\Theta}\boldsymbol{f}_{m,j}|^2},\tag{28}\label{28}
\end{align}
Then, the optimization for passive beamforming at RIS can be formulated as
%%%%%%%%%%%%%%%%%%%%%%%%%%%
\begin{alignat}{2}
&\underset{\boldsymbol{\Theta}}{\text{max}}\ \log_2(1+\bar\gamma_{l,i})+\log_2(1+\bar\gamma_{l,j})\tag{29}\label{OP2} \\
 s.t.\  &    \bar\gamma_{l,i}\geq \bar\gamma_{min}, \tag{29a}\label{29a}\\
 & \bar\gamma_{l,j}\geq \bar\gamma_{min},\tag{29b}\label{29b} \\
   & |\alpha_m|=1, m\in M, \tag{29c}\label{29c}
\end{alignat}
%%%%%%%%%%%%%%%%%%%%%%%%%%
 To easily handle the terms $|\boldsymbol{g}_{l,m}\boldsymbol{\Theta}\boldsymbol{f}_{m,i}|^2$ and $|\boldsymbol{g}_{l,m}\boldsymbol{\Theta}\boldsymbol{f}_{m,j}|^2$, let us denote $\boldsymbol{\xi}=[\xi_1,\xi_2,\xi_3,\dots,\xi_{m}]$ is the diagonal vector of reconfigurable elements in passive beamforming matrix $\boldsymbol{\Theta}$, where $\xi_m=\alpha^H_m$. Then, we introduce an auxiliary vector $\boldsymbol{\hat{H}}_{l,\iota}$ such that $\boldsymbol{\hat{H}}_{l,\iota}=\boldsymbol{g}_{l,m}\circ\boldsymbol{f}_{m,\iota}$, where $\circ$ represents the Hadamard product and $\iota\in\{i,j\}$. Now the following equality can efficiently hold 
as $|\boldsymbol{g}_{l,m}\boldsymbol{\Theta}\boldsymbol{f}_{m,\iota}|^2=|\boldsymbol{\xi}^H\boldsymbol{\hat{H}}_{l,\iota}|^2$. Adopting these updates, the problem in (\ref{OP2}) can be rewritten as
 %%%%%%%%%%%%%%%%%%%%%%%%%%%
\begin{alignat}{2}
&\underset{\boldsymbol{\xi}}{\text{max}}  \ \log_2(1+P_l\varrho_{l,i}|\boldsymbol{\xi}^H\boldsymbol{\hat{H}}_{l,i}|^2/\sigma^2)\nonumber\\&+\log_2\bigg(1+\frac{P_l\varrho_{l,j}|\boldsymbol{\xi}^H\boldsymbol{\hat{H}}_{l,j}|^2}{P_l\varrho_{l,i}|\boldsymbol{\xi}^H\boldsymbol{\hat{H}}_{l,j}|^2+\sigma^2}\bigg)\tag{30}\label{OP21} \\
 s.t.\  &   P_l\varrho_{l,i}|\boldsymbol{\xi}^H\boldsymbol{\hat{H}}_{l,i}|^2\geq \bar\gamma_{min}(\sigma^2), \tag{30a}\label{30a}\\
 & P_l\varrho_{l,j}|\boldsymbol{\xi}^H\boldsymbol{\hat{H}}_{l,j}|^2\geq \bar\gamma_{min}(P_l\varrho_{l,i}|\boldsymbol{\xi}^H\boldsymbol{\hat{H}}_{l,j}|^2+\sigma^2), \tag{30b}\label{30b} \\
   & |\xi_m|=1, m\in M,\tag{30c}\label{30c}
\end{alignat}
%%%%%%%%%%%%%%%%%%%%%%%%%%
To solve (\ref{OP21}), we re-expressed $P_l\varrho_{l,\iota}|\boldsymbol{\xi}^H\boldsymbol{\hat{H}}_{l,\iota}|^2$ as
\begin{align}
& P_l\varrho_{l,\iota}|\boldsymbol{\xi}^H\boldsymbol{\hat{H}}_{l,\iota}|^2= \boldsymbol{\xi}^HP_l\varrho_{l,\iota}\boldsymbol{\hat{H}}_{l,\iota}\boldsymbol{\hat{H}}^H_{l,\iota}\boldsymbol{\xi}\nonumber\\& = \text{Tr}(\boldsymbol{\xi}^HP_l\varrho_{l,\iota}\boldsymbol{\hat{H}}_{l,\iota}\boldsymbol{\hat{H}}^H_{l,\iota}\boldsymbol{\xi})=\text{Tr}(P_l\varrho_{l,\iota}\boldsymbol{\hat{H}}_{l,\iota}\boldsymbol{\hat{H}}^H_{l,\iota}\boldsymbol{\xi}\boldsymbol{\xi}^H).\tag{31}
\end{align}
Next, we define two auxiliary matrices as $\boldsymbol{G}_{l,\iota}=P_l\varrho_{l,\iota}\boldsymbol{\hat{H}}_{l,\iota}\boldsymbol{\hat{H}}^H_{l,\iota}$ and $\boldsymbol{\Xi}=\boldsymbol{\xi}\boldsymbol{\xi}^H$. We can observe that $\boldsymbol{G}_{l,\iota}$ and $\boldsymbol{\Xi}$ are semi-definite. Now the problem in (\ref{OP21}) can be reformulated as
 %%%%%%%%%%%%%%%%%%%%%%%%%%%
\begin{alignat}{2}
&\underset{\boldsymbol{\Xi}}{\text{max}} \ \log_2(1+\text{Tr}(\boldsymbol{\Xi}\boldsymbol{G}_{l,i})/\sigma^2)\nonumber\\&+\log_2\bigg(1+\frac{\text{Tr}(\boldsymbol{\Xi}\boldsymbol{G}_{l,j})}{\text{Tr}(\boldsymbol{\Xi}\boldsymbol{\bar{G}}_{l,j})+\sigma^2}\bigg)\tag{32}\label{OP22} 
\end{alignat}
\begin{align}
 s.t.\  &  \text{Tr}(\boldsymbol{\Xi}\boldsymbol{G}_{l,i})\geq \bar\gamma_{min}(\sigma^2), \tag{32a}\label{32a}
 \end{align}
 \begin{align}
 & \text{Tr}(\boldsymbol{\Xi}\boldsymbol{G}_{l,j})\geq \bar\gamma_{min}(\text{Tr}(\boldsymbol{\Xi}\boldsymbol{\bar{G}}_{l,j})+\sigma^2), \tag{32b}\label{32b} 
 \end{align}
\begin{align}
   & \text{diag}\{\boldsymbol{\Xi}\}=1, \tag{32c} \label{32c}
    \end{align}
\begin{align}
& \boldsymbol{\Xi}\succeq 0, \tag{32d}\label{32d}
 \end{align}
\begin{align}
& \text{rank}(\boldsymbol{\Xi}) = 1, \tag{32e}\label{32e}
\end{align}
%%%%%%%%%%%%%%%%%%%%%%%%%%
where $\boldsymbol{\bar{G}}_{l,j}=P_l\varrho_{l,i}\boldsymbol{\hat{H}}_{l,j}\boldsymbol{\hat{H}}^H_{l,j}$. The $\text{diag}\{\boldsymbol{\Xi}\}$ is used to return the diagonal values of $\boldsymbol{\Xi}$ which means $\text{diag}\{\boldsymbol{\Xi}\}=1$ equals to $\xi^2_m=1$ and $|\xi_m|=1$.

The optimization in (\ref{OP22}) is still non-convex due to the objective function and the rank one constraint. Let us start with the rank one constraint in (\ref{32e}), which can be efficiently replaced by a semi-definite (convex) constraint as $\boldsymbol{\Xi}-\boldsymbol{\hat{\xi}}\boldsymbol{\hat{\xi}}^H\succeq 0$, where $\boldsymbol{\hat{\xi}}$ denotes a vector of auxiliary variables. Now we can replaced $\boldsymbol{\Xi}-\boldsymbol{\hat{\xi}}\boldsymbol{\hat{\xi}}^H\succeq 0$ by convex Schur complement as
\begin{align}
\begin{bmatrix}
\boldsymbol{\Xi} & \boldsymbol{\hat{\xi}} \\
\boldsymbol{\hat{\xi}}^H & 1
\end{bmatrix} \succeq 0.\tag{33}
\end{align}
Next, we take the objective function of the problem (\ref{OP22}), which can be efficiently written as Equation (\ref{253}), provided on the top of the next page.
It can be observed that (\ref{253}) is a DC function and can be handled by DC programming. However, the $\boldsymbol{\Xi}$ is a complex matrix and calculating its partial derivation through the traditional way is impossible. Thus, it is necessary to calculate its derivation for both imaginary and real parts. After incorporating the above changes, the problem (\ref{OP22}) becomes semi-definite programming, which is convex. This problem can be efficiently handled through any standard convex solver such as CVX.

\subsection{Proposed Algorithm and Complexity Analysis}
This work has provided an energy efficiency optimization framework in RIS-assisted NOMA LEO satellite communication networks. In particular, the proposed solution has been achieved simultaneously in two steps. In the first step, the power of the LEO satellite for GUs is calculated for any given phase shift design at the RIS system. Then, in the next step, passive beamforming is designed at RIS, given the LEO satellite power. The detailed process of the proposed solution is also given in Algorithm 1. It is important to discuss the complexity of the proposed solution. In this work, the complexity can be defined as the number of iterations required for the convergence of optimization variables. The complexity of the power allocation solution in the first step in a single iteration can be described as $\mathcal O\{2(i+j)\}$, where $i$ and $j$ are the NOMA GUs. For passive beamforming in the second step, the complexity can be expressed as $\mathcal O(M^{3.5})$ \cite{khan2022integration}. Let us assume that $T$ is the number of iterations which is required for convergence. Then, the total complexity of the proposed solution involving the first and second steps can be stated as $\mathcal O[T(2(i+j)+M^{3.5})]$.
%%%%%%%%%%%%%%%%%%%%%%%%%%%%%%%%%%%%%%%%%%
\begin{algorithm}[t]
{\bf Initialize:} we first initialize all the system parameters and set iteration index $t=1$
 
{\bf Step 1:} calculate $\varrho_{l,i}$ and $\varrho_{l,j}$ for the given $\boldsymbol{\Theta}$
         
 \While{not converge}{\For{$t=1:t_{max}$}{Compute $\frac{R_{l,i}+R_{l,j}}{P_l(\varrho_{l,i}+\varrho_{l,j})+p_c}(t)$ \\
Update ${\lambda_1,\lambda_2,\lambda_3,\lambda_4}$ by (\ref{23})-(\ref{26})\\
Compute $\varrho_{l,i}$ and $\varrho_{l,j}$ using (\ref{18}) and (\ref{22}) }
{\bf Step 2:} Having $\varrho^*_{l,i}$ and $\varrho^*_{l,j}$, we compute $\boldsymbol{\Theta}$\\
\For{$t=1:t_{max}$}{1) Transformation of problem in (29) into convex semi-definite programming problem\\
2) Then, use a standard convex solver such as CVX to compute passive beamforming}}
Return $\varrho^*_{l,i}$, $\varrho^*_{l,j}$, $\boldsymbol{\Theta^*}$
\caption{Energy Efficiency Optimization in RIS-Assisted NOMA Satellite Communication Network}
\end{algorithm}  
   %%%%%%%%%%%%%%%%%%%%%%%%%%%%%%%%%%%%%%%%%%
%%%%%%%%%%%%%%%%
\begin{figure}[!t]
\centering
\includegraphics [width=0.50\textwidth]{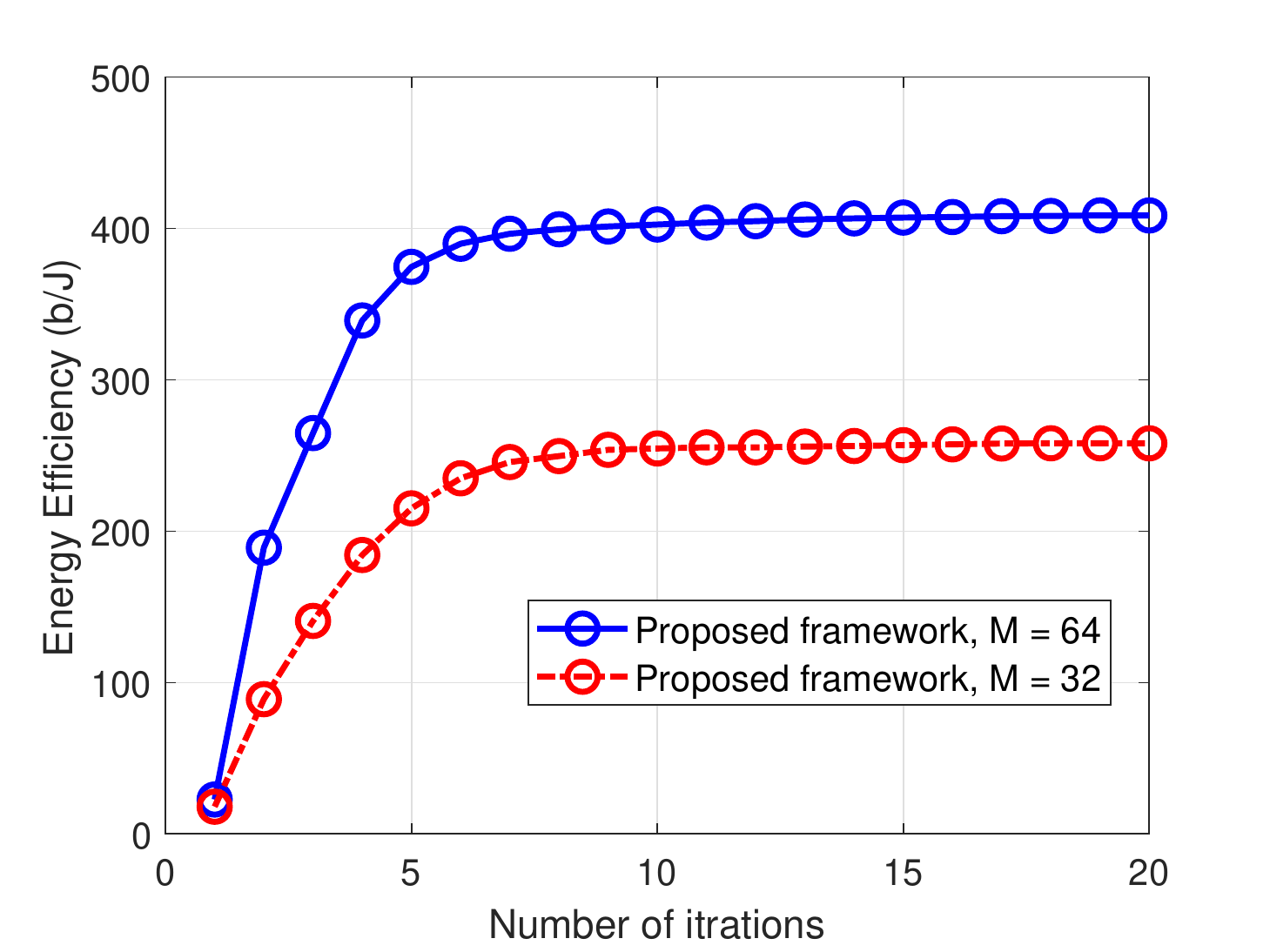}
\caption{Convergence of the proposed optimization framework.}
\label{Fig1VTC}
\end{figure}
%%%%%%%%%%%%%%%%%%%%%%%%%%%%%
\section{Results and Discussion}
In this section, we present the numerical results obtained through Monte Carlo simulations. We compare three frameworks: the proposed framework (Section III), the benchmark framework (where the phase shift vector of RIS is fixed and only the transmit power of the LEO satellite is optimized), and the conventional framework (where RIS is not involved in the proposed model). Unless mentioned otherwise, the simulation parameters are set as follows \cite{8911361,8786872,9583591,10097680}. We consider that the proposed RIS-assisted NOMA LEO satellite network is operating over Ka-band (17.7-19.7 GHz). The number of NOMA GUs is 2, the number of reconfigurable elements at RIS is 64, the maximum transmission power of the LEO satellite is 50 dBm, the quality of services per GU is set as 10 b/s, and the bandwidth of each GU is 20 MHz. Moreover, noise power density is set as -170 dBm, $G_{l,\iota}=3.5$ dBi, $G_{l}=20$ dBi, $G_{max}=52.1$ dBi, and the height of the LEO satellite is 554.8 km \cite{9420293}.

Fig. \ref{Fig1VTC} illustrates the convergence behaviour of the proposed optimization framework. The plot shows the achievable energy efficiency as a function of the number of iterations for the systems with 34 and 64 reconfigurable RIS elements. The results demonstrate that the proposed framework achieves convergence after only 8 iterations, indicating its low-complexity nature. Furthermore, it can be observed that the system with more reconfigurable RIS elements achieves higher energy efficiency compared to the system with fewer reconfigurable elements. Additionally, the convergence behaviour of the proposed framework for both systems is similar, highlighting its suitability for large-scale systems.

%%%%%%%%%%%%%%%%%%%%%%%%%%%%%
\begin{figure}[!t]
\centering
\includegraphics [width=0.50\textwidth]{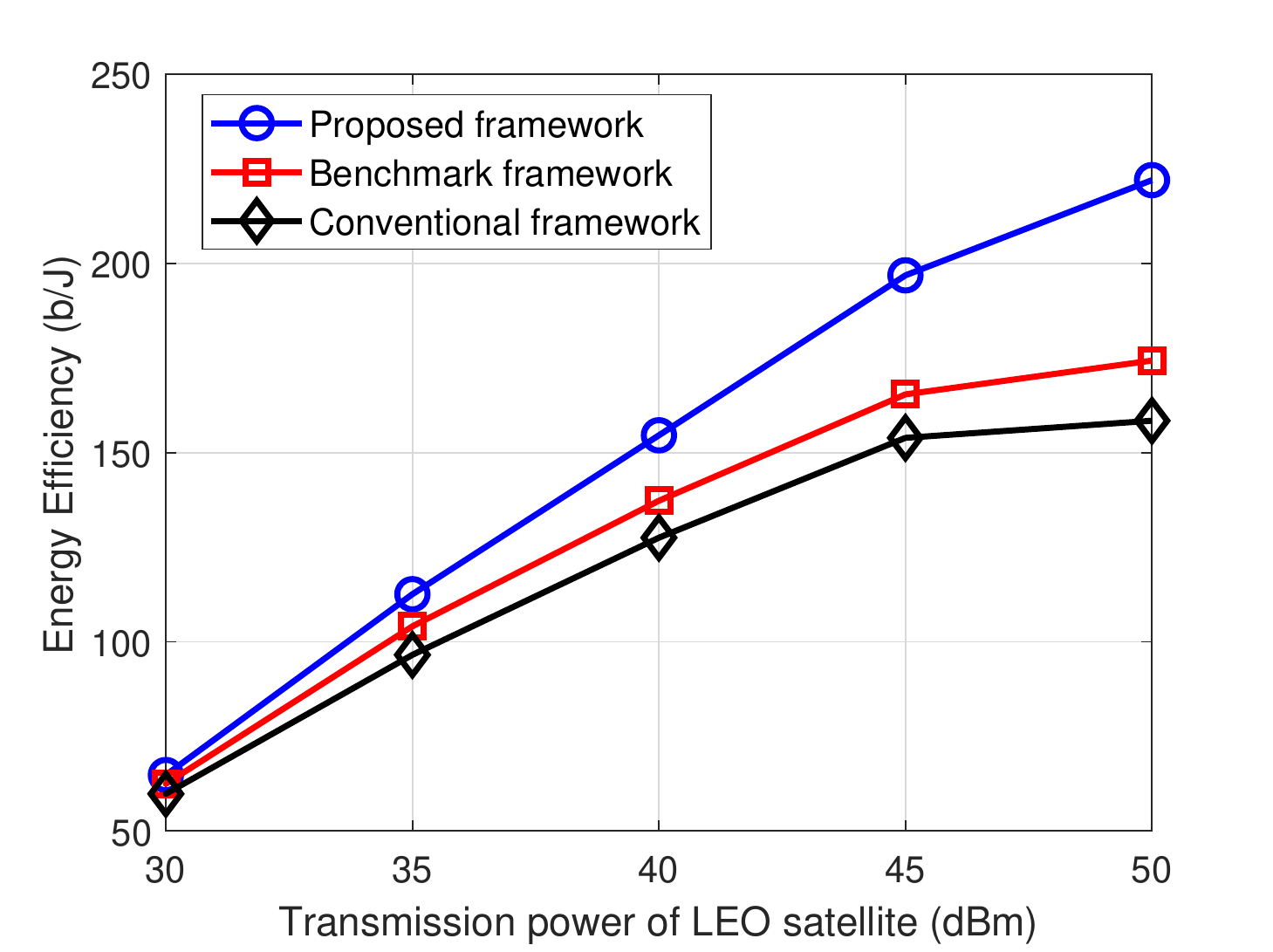}
\caption{Available transmit power versus energy efficiency of the LEO satellite network for $M=32$ and $\gamma_{min}=10$ b/s.}
\label{Fig2VTC}
\end{figure}
%%%%%%%%%%%%%%%%%%%%%%%%%%%%%

Figure \ref{Fig2VTC} illustrates the impact of LEO power on the achievable energy efficiency of the satellite network for all three frameworks. The results demonstrate that the achievable energy efficiency of the system increases for all three frameworks as the available transmit power increases. Additionally, we can see that the results of all three frameworks follow a bell-shaped curve, where the achievable energy efficiency exponentially increases with an increase in the transmission power of the LEO satellite and then slows down towards saturation. However, the proposed framework consistently outperforms the other two frameworks for all values of transmit power. Moreover, the gap between the proposed framework and the other frameworks increases as the transmit power increases, indicating the poor performance of the benchmark and conventional frameworks.

%%%%%%%%%%%%%%%%%%%%%%%%%%%%%
\begin{figure}[!t]
\centering
\includegraphics [width=0.51\textwidth]{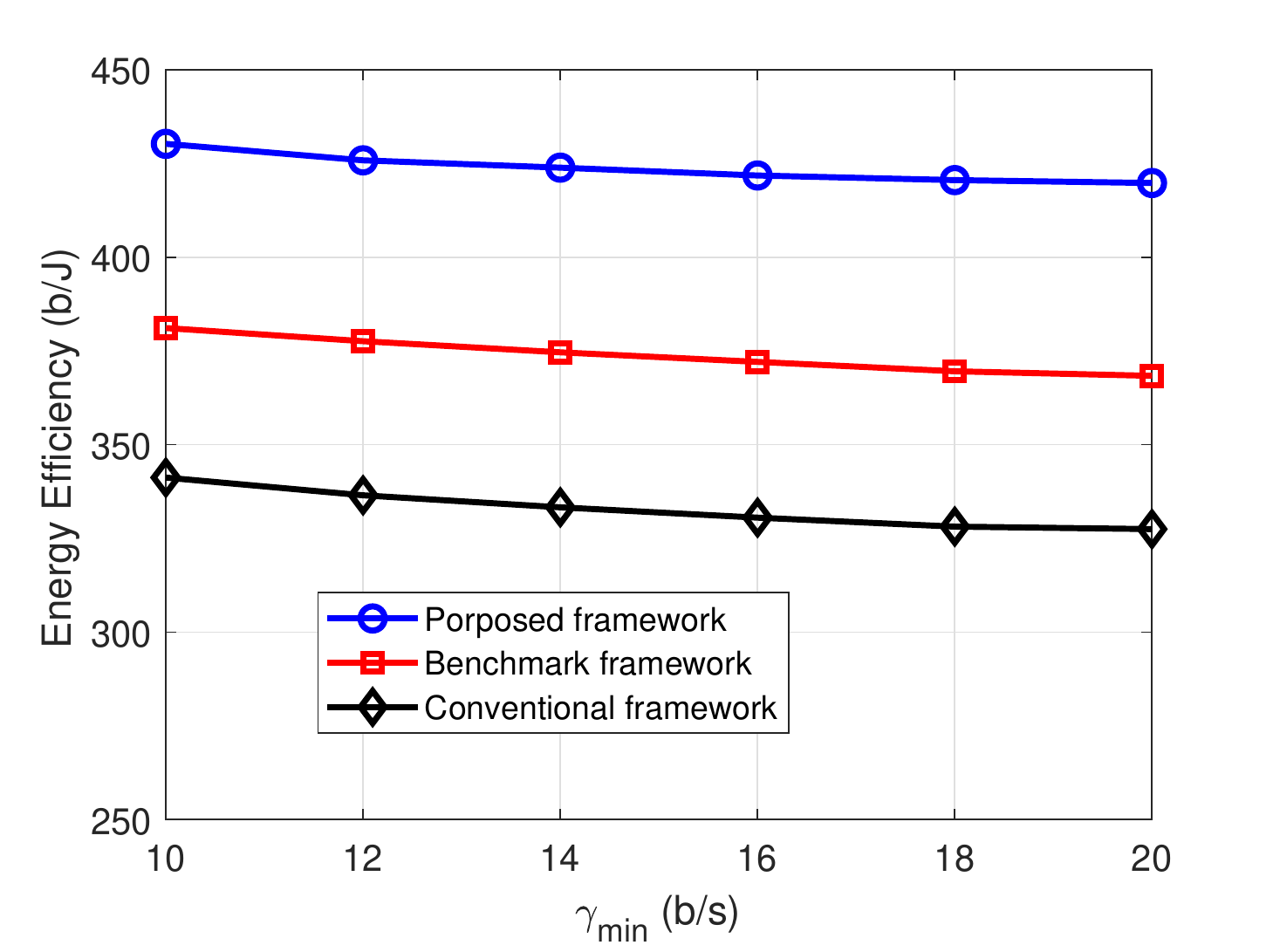}
\caption{Quality of services requirements versus achievable energy efficiency of the LEO satellite network for $P_l=50$ and $M=64$.}
\label{Fig3VTC}
\end{figure}
%%%%%%%%%%%%%

Fig. \ref{Fig3VTC} shows the achievable energy efficiency of the system against the increasing values of quality of services requirements. The results show that the achievable energy efficiency of the system slightly decreases for all frameworks as the transmit power of the LEO satellite increases. This is due to the fact that, as the value of $\gamma_{min}$ increases, more power is required for GUs to meet the minimum quality of service requirements, thereby affecting the achievable energy efficiency of the LEO satellite network. However, the proposed framework achieves higher energy efficiency compared to the other two frameworks. Additionally, it can be observed that the benchmark framework performs better than the conventional framework, highlighting the potential benefits of using RIS in satellite networks. 
%%%%%%%%%%%%%%%%%%%%%%%%%%%%%
\begin{figure}[!t]
\centering
\includegraphics [width=0.50\textwidth]{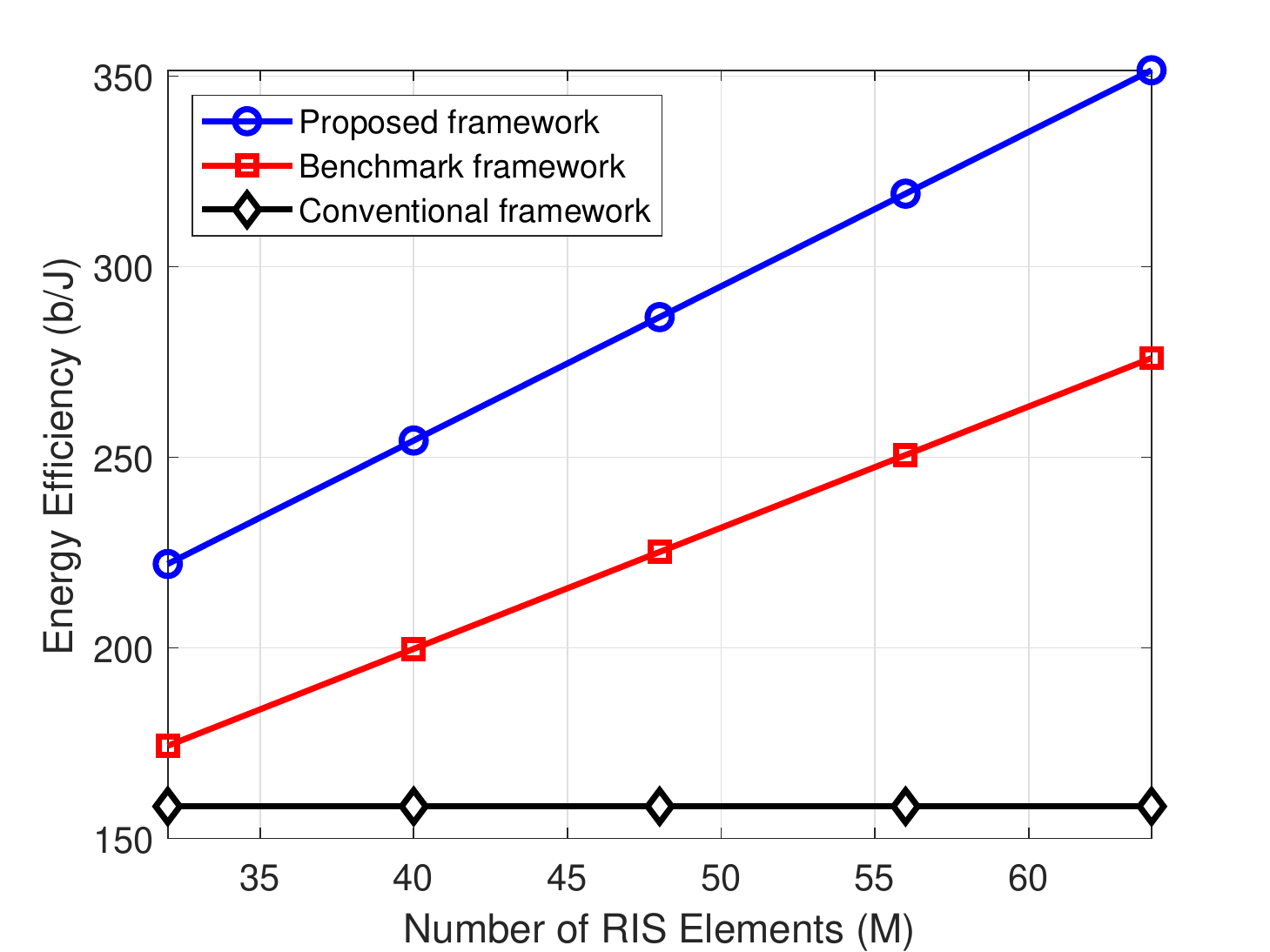}
\caption{Number of reconfigurable elements at RIS versus achievable energy efficiency of the LEO satellite network for $P_l=50$ and $\gamma_{min}=20$ b/s.}
\label{Fig4VTCext}
\end{figure}
%%%%%%%%%%%%%

Fig. \ref{Fig4VTCext} demonstrates the achievable energy efficiency of the system against the number of RIS reconfigurable elements. We can see that the achievable energy efficiency of the proposed framework and benchmark framework increases as the number of RIS elements increases. However, the proposed framework with optimal beamforming achieves high energy efficiency compared to the benchmark framework, which follows a fixed phase shift policy. For example, if we fix the transmission power at 50 dBm and vary the number of elements at RIS from 32 to 64. The achievable energy efficiency of the proposed framework increases from 222 to 531.5 b/J. However, on the same system parameters, the benchmark framework energy efficiency is well behind and has improved from 174.3 to 276 b/J. Moreover, the conventional framework, which does not involve the RIS system, remains unchanged at 158.4 b/J.

\begin{figure}[!t]
\centering
\includegraphics [width=0.50\textwidth]{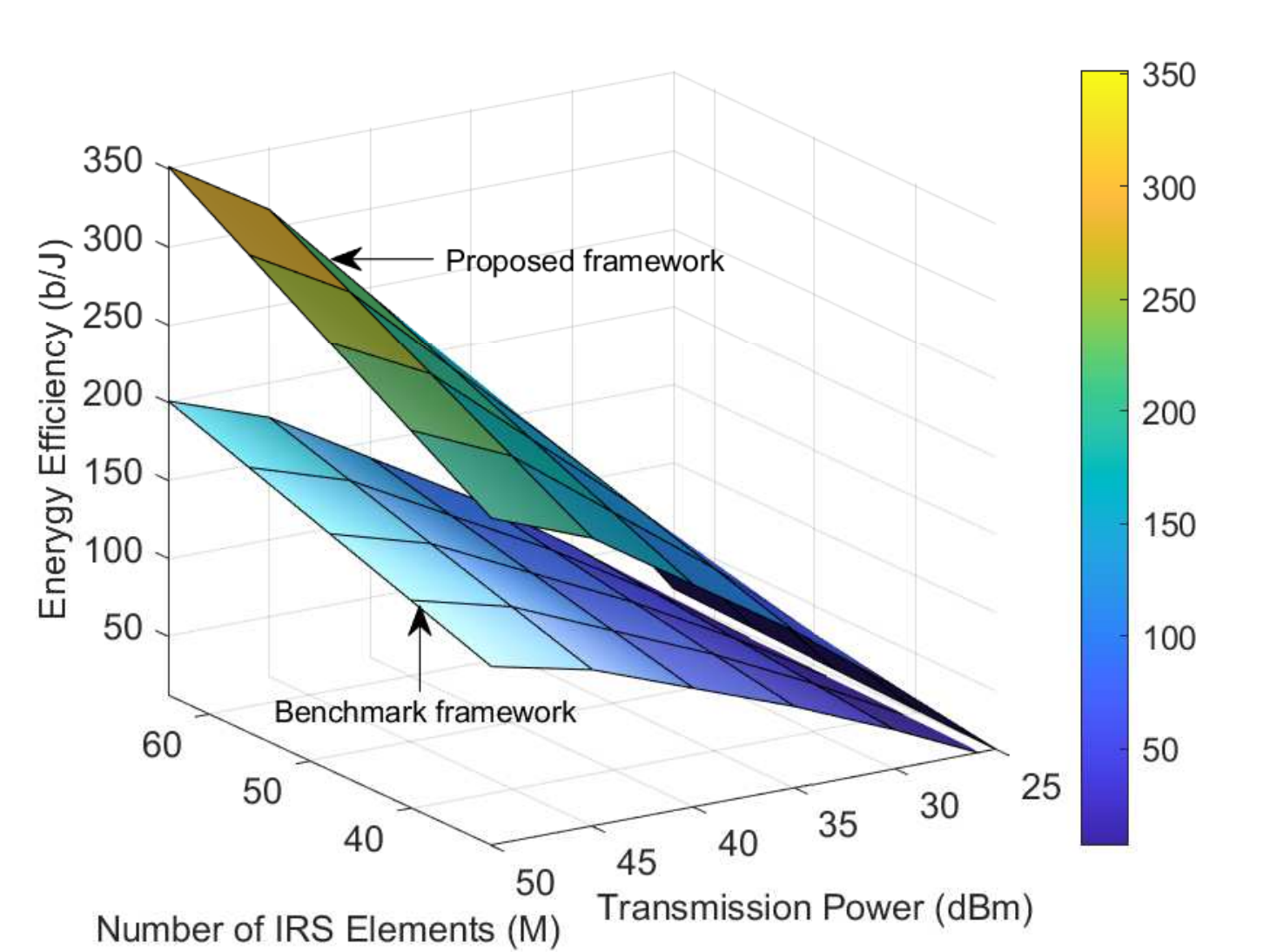}
\caption{Impact of transmission power and reconfigurable elements on the achievable energy efficiency of the LEO satellite network for $\gamma_{min}=20$ b/s.}
\label{Fig5VTCext}
\end{figure}

Fig. \ref{Fig5VTCext} shows the impact of the number of reconfigurable elements at RIS and the transmission power of the LEO satellite on the system's achievable energy efficiency. The figure demonstrates that the transmission power of the LEO satellite has a greater influence on the performance of the RIS, consequently affecting the system's achievable energy efficiency. For instance, when keeping the transmission power of the LEO satellite fixed at 30 dBm and varying the number of reconfigurable elements at the RIS from 32 to 64, the proposed framework achieves an increase in energy efficiency from approximately 64.7 to 102.5 b/J. However, if we fix the transmission power of LEO at 50 dBm and increase the number of reconfigurable elements at RIS from 32 to 64, the system's achievable energy efficiency of the proposed framework increases from 222 to 351.5 b/J. Moreover, it can also be noted that for the same system parameters, the benchmark framework achieves lower energy efficiency compared to the proposed framework.

%%%%%%%%%%%%%%%%%%%%%%%%%%%%%
%%%%%%%%%%%%%
\section{Conclusions}
In this paper, we present a new framework for optimizing the energy efficiency of RIS-assisted NOMA LEO satellite communication networks. The transmit power of the LEO satellite and the passive beamforming of RIS are simultaneously optimized to maximize the achievable energy efficiency of the satellite network while ensuring the quality of services. The non-convex optimization problem is addressed through alternating optimization in two steps. Firstly, the NOMA power allocation at the LEO satellite is computed by successive convex optimization and dual Lagrangian methods, given the fixed phase shift at RIS. Then, passive beamforming at RIS, a standard convex method, is adopted to handle the semi-definite programming, given the power allocation at the LEO satellite. Numerical results demonstrate that the proposed framework achieves high energy efficiency and converges within a few iterations.

\bibliographystyle{IEEEtran}
\bibliography{Wali_EE}

\end{document}